\documentclass[12pt,onecolumn,draftclsnofoot]{IEEEtran}
\usepackage{ifpdf}
\usepackage{cite}
\ifCLASSINFOpdf
   \usepackage[pdftex]{graphicx}
\else
   \usepackage[dvips]{graphicx}
\fi
\usepackage{amsmath}
\usepackage{algorithmic}
\usepackage{array}
\ifCLASSOPTIONcompsoc
  \usepackage[caption=false,font=normalsize,labelfont=sf,textfont=sf]{subfig}
\else
  \usepackage[caption=false,font=footnotesize]{subfig}
\fi
\usepackage{fixltx2e}

\usepackage{bbm}
\usepackage{amssymb}
\usepackage{amscd}
\usepackage[amsmath,thref,thmmarks]{ntheorem}
\theoremstyle{plain}
\newtheorem{proposition}{Proposition}
\newtheorem{theorem}{Theorem}
\newtheorem{definition}{Definition}
\newtheorem{lemma}{Lemma}

\usepackage{stmaryrd}

\usepackage{hyperref}

\begin{document}
\title{Secure Polar Coding for Adversarial Wiretap Channel}

\author{Yizhi~Zhao
%~and~Hongmei~Chi% <-this % stops a space
\thanks{Y. Zhao was with the College of Informatics, Huazhong Agricultural University, Wuhan,
Hubei, China. E-mail: zhaoyz@mail.hzau.edu.cn.}% <-this % stops a space
%\thanks{H. Chi  was with the College of Science, Huazhong Agricultural University, Wuhan,
%Hubei, China. E-mail: chihongmei@mail.hzau.edu.cn.}% <-this % stops a space
}

\maketitle

\begin{abstract}
The adversarial wiretap channel (AWTC) model is a secure communication model in which adversary can directly read and write the transmitted bits in legitimate communication with fixed fractions. In this paper we propose a secure polar coding scheme to provide secure and reliable communication over the AWTC model. For the adversarial reading and writing action, we present a $\rho$ equivalent channel block and study its transformation under the channel polarization operation. We find that the generated channels are polarized in the sense that part of them are full-noise channels with probability almost $1$ and the rest part of them are noiseless channels with probability almost $1$. Based on this result, we polarize both equivalent channel blocks of adversarial reading and writing, and then construct a secure polar coding scheme by applying the multi-block chaining structure on the polarized equivalent blocks. Theoretically we prove that when block length $N$ goes infinity, the proposed scheme achieves the secrecy capacity of the AWTC model under both reliability and strong security criterions. Then by simulations, we prove that the proposed scheme can provide secure and reliable communication over AWTC model.
\end{abstract}


% \fi
\IEEEpeerreviewmaketitle

\section{Introduction}

\IEEEPARstart{W}iretap channel (WTC) model is a primitive secure communication model introduced by Wyner in 1975\cite{Wyner1975}. In the WTC model, two legitimate users, Alice and Bob, communicate through a noise main channel while an eavesdropper Eve is wiretapping through a noise wiretap channel. Then in \cite{Ozarow1984}, the wiretap channel model type II (WTC-II) was introduced, in which the main channel between legitimate parties becomes noiseless, but the eavesdropper Eve can directly read a fixed fraction $\alpha$ of transmitted bits. Further in \cite{Wang2015}, as an extension of the WTC-II model, the adversarial wiretap channel (AWTC) model was introduced, in which the adversary Eve can both read and write the transmitted bits with fixed fractions $(\rho_r,\rho_w)$.

According to Wyner's physical layer secure coding theory\cite{Wyner1975}, secure codes can make use of the channel noise to provide perfect secrecy over WTC models and achieve the secrecy capacities. For Wyner's coding theory, one effective coding method is the polar codes which is invented by Ar{\i}kan and known as the first capacity achieving code with low complexity \cite{Arikan2009}. Over the last decades, explicit secure polar codes successfully constructed and achieved the secrecy capacities for Wyner's WTC model\cite{Andersson2010,Hof2010,Mahdavifar2011,Vard2013strong} and several extended WTC models \cite{Andersson2013,Wei2015,Chou2016,Si2016,Gulcu2017,Chou2018,Chou2018_2,Chou2020}. These success of secure polar codes on WTC models make polar codes a considerable good option for achieving the secrecy capacity of AWTC model.

Unlike the WTC model that both main channel and wiretap channel are known and fixed, AWTC model is more complicated when constructing the secure polar codes. In the AWTC model, adversary's reading and writing operations act directly on the transmitted information, so there is no ``noisy channels" for constructing the secure polar codes. Moreover, the indices of the transmitted bits for each reading and writing operation are randomly selected by the adversary and inaccessible to legitimate parties, making it more difficult to construct the secure polar codes.

\subsection{Our Contributions}

In this work, we investigate the construction of explicit secure polar codes for AWTC model to achieve reliability, security and the secrecy capacity. Our contributions are summarized as follow:
\begin{itemize}
\item For adversary's reading and writing operations, we present an equivalent model as the \emph{$\rho$ equivalent channel block} $W_\rho^{1:N}$ (Def.~\ref{def_rec}) which is an $N$ length non-stationary channel block composed of full-noise BECs with fraction $\rho$ and noiseless BECs with fraction $1-\rho$. For reading operation with fraction $\rho_r$, the equivalent channel block is $W_{1-\rho_r}^{1:N}$. For writing operation with fraction $\rho_w$, the equivalent channel block is $W_{\rho_w}^{1:N}$.
\item Then we apply the channel polarization operation $\mathbf{G}_N$ on the $\rho$ equivalent channel block and prove that the generated channels $W_{\rho,N}^{(1:N)}$ is polarized in the sense that, for any fixed $\delta_N=2^{-N^\beta}$, $\beta\in(0,\frac{1}{2})$, as $N=2^n$ goes to infinity, the fraction of indices $i\in[\![1,N]\!]$ for which $P(W_{\rho,N}^{(i)})\in[0,\delta_N)$ goes to $1-\rho$ and the fraction of indices $i\in[\![1,N]\!]$ for which $P(W_{\rho,N}^{(i)})\in(1-\delta_N,1]$ goes to $\rho$, where $P(W_{\rho,N}^{(i)})\triangleq\Pr\{Z(W_{\rho,N}^{(i)})=1\}$ (Prop.~\ref{prop_polarization}).
\item According to the polarization result of $\rho$ equivalent channel block, we polarize the writing equivalent block $W_{\rho_w}^{1:N}$ and the reading equivalent block $W_{1-\rho_r}^{1:N}$. Based on the polarization results we apply the multi-block chaining structure\cite{Vard2013strong} to construct a strong security polar coding scheme. Theoretically, we have proven that the proposed scheme achieves reliability, strong security and secrecy capacity when the block length $N$ goes infinity. We also carry out simulations to test the performance of the proposed scheme. The simulation results match our theoretically analysis and further prove that the proposed scheme successfully provide a reliable and secure communication over AWTC model.
\end{itemize}

\subsection{Related Works}

The first secrecy capacity achieving secure polar codes for WTC model was proposed in \cite{Mahdavifar2011} which sets up a basic way of secure polar codes construction that use the differences of polarizations between main channel block and wiretap channel block to divide the channel index into subsets with unique reliability and security properties. Further in \cite{Vard2013strong}, a multi-block chaining structure was constructed as a refinement of secure polar coding scheme in \cite{Mahdavifar2011}, which is known as one of the standard method for strong security polar coding construction and widely applied in the follow-up works such as \cite{Wei2015,Gulcu2017}. One remarkable advantage of this standard strong security polar coding method is the low computational complexity for both encoding and decoding process as $O(N\log N)$.

When the AWTC model was proposed in \cite{Wang2015}, an effective explicit secrecy capacity achieving codes named as the \emph{capacity achieving AWTC code family} was also proposed which contains three building blocks: algebraic manipulation detection code (AMD code), subspace evasive sets, and folded reed-solomon code (FRS code). For capacity achieving AWTC code family in \cite{Wang2015}, the encoding complexity is $O((N\log q)^2)$ where $q$ is a prime satisfying $q>Nu$ for a $u$-folded RS codes, the total complexity of the decoding is $\mathrm{poly}(N)$. Then for our proposed secure polar coding scheme, the polarization process of the $\rho$ equivalent channel block is the same as the polarization of regular BEC block, and we only apply the standard method of strong security polar coding construction, thus both encoding and decoding computational complexities are $O(N\log N)$, which is lower than the secure codes in \cite{Wang2015}.

In \cite{Chou2018_2} and \cite{Chou2020}, the author proposed a unified design framework of explicit secure polar codes for multiple WTC models, including Wyner's WTC model, WTC-II model, WTC model with compound eavesdropper channel and WTC model with arbitrarily varying eavesdropper channel. By using the polar code based random binning technique, this unified design framework achieves reliability, strong security and secrecy capacities over those WTC models. However this unified design framework is not applicable to the AWTC model since the main channel needs to be known and fixed.

\subsection{Paper Organizations}

The rest of this paper is organized as follow. Section \ref{sec_awtc} presents the AWTC model. Section \ref{sec_polarization} presents the equivalent model of AWTC and its polarization discussion. Section \ref{sec_code} presents our construction of secure polar coding scheme for AWTC model and the performance analysis. Finally section \ref{sec_con} concludes the paper.

%\IEEEpubidadjcol

\section{The Adversarial Wiretap Channel}\label{sec_awtc}

\emph{Notations:} We define the integer interval $[\![a,b]\!]$ as the integer set between $\lfloor a\rfloor$ and $\lceil b\rceil$. For $n\in \mathbb{N}$, define $N\triangleq 2^n$. Denote $X$, $Y$, $Z$,... random variables (RVs) taking values in alphabets $\mathcal{X}$, $\mathcal{Y}$, $\mathcal{Z}$,... and the sample values of these RVs are denoted by $x$, $y$, $z$,... respectively. Then $p_{XY}$ denotes the joint probability of $X$ and $Y$, and $p_X$, $p_Y$ denotes the marginal probabilities. Especially for channel $W$, the transition probability is defined as $W_{Y|X}$ and $W$ for simplicity. Also we denote a $N$ size vector $X^{1:N}\triangleq (X^1,X^2,...,X^N)$. When the context makes clear that we are dealing with vectors, we write $X^N$ in place of $X^{1:N}$. And for any index set $\mathcal{A}\subseteqq [\![1,N]\!]$, we define $X^{\mathcal{A}}\triangleq \{X^i\}_{i\in \mathcal{A}}$. For the polar codes, we denote $\mathbf{G}_N$ the generator matrix of polar codes, $\mathbf{R}$ the bit reverse matrix, $\mathbf{F}=
    \begin{bmatrix}\begin{smallmatrix}
        1 & 0 \\
        1 & 1
    \end{smallmatrix}\end{bmatrix}$
and $\otimes$ the Kronecker product, and we have $\mathbf{G}_N=\mathbf{R}\mathbf{F}^{\otimes n}$.

Now we present the definition of adversarial wiretap channel model (AWTC)\cite{Wang2015}.

\begin{definition}\label{def_awt} The adversarial wiretap channel model is defined as $(\mathcal{X},\mathcal{Y},\mathcal{Z}, \rho_r,\rho_w,\mathcal{S}_r,\mathcal{S}_w)$. In the model, legitimate parties are communicating through a noiseless channel with channel input alphabet $\mathcal{X}$. For $N$-length transmitted codewords $X^{1:N}\in \mathcal{X}^N$, there are two types of adversarial actions: reading and writing.
\begin{itemize}
\item \textbf{Reading:} Eavesdropper can arbitrarily select an index subset $\mathcal{S}_r\subseteq[\![1,N]\!]$ with fixed fraction $\rho_r=\frac{|\mathcal{S}_r|}{N}$ and directly reads the corresponding transmitted codewords $X^{\mathcal{S}_r}$. For the obtained bits at eavesdropper, $Z^N$ with alphabet $\mathcal{Z}$, has
    \begin{equation}
    Z^i=
    \begin{cases}
    X^i&\text{~if~}i\in\mathcal{S}_r\\
    ?&\text{~if~}i\in\mathcal{S}_r^c
    \end{cases}
    \end{equation}
    where ``$?$" is the dump letter.
\item \textbf{Writing:} Eavesdropper can arbitrarily select an index subset $\mathcal{S}_w\subseteq[\![1,N]\!]$ with fixed fraction $\rho_w=\frac{|\mathcal{S}_w|}{N}$ and directly writes the corresponding transmitted codewords $X^{\mathcal{S}_w}$. For the channel output at legitimate receiver, $Y^N$ with alphabet $\mathcal{Y}$, has
    \begin{equation}
    Y^i=
    \begin{cases}
    ? &\text{~if~}i\in\mathcal{S}_w\\
    X^i &\text{~if~}i\in\mathcal{S}_w^c
    \end{cases}
    \end{equation}
    where ``$?$" is the dump letter.
    \end{itemize}

Both adversarial reading and writing act independently on the transmitted bit $X^N$, but we assume that $\rho_w+\rho_r<1$.
\end{definition}

\begin{figure}[!h]
\centering
\includegraphics[width=13cm]{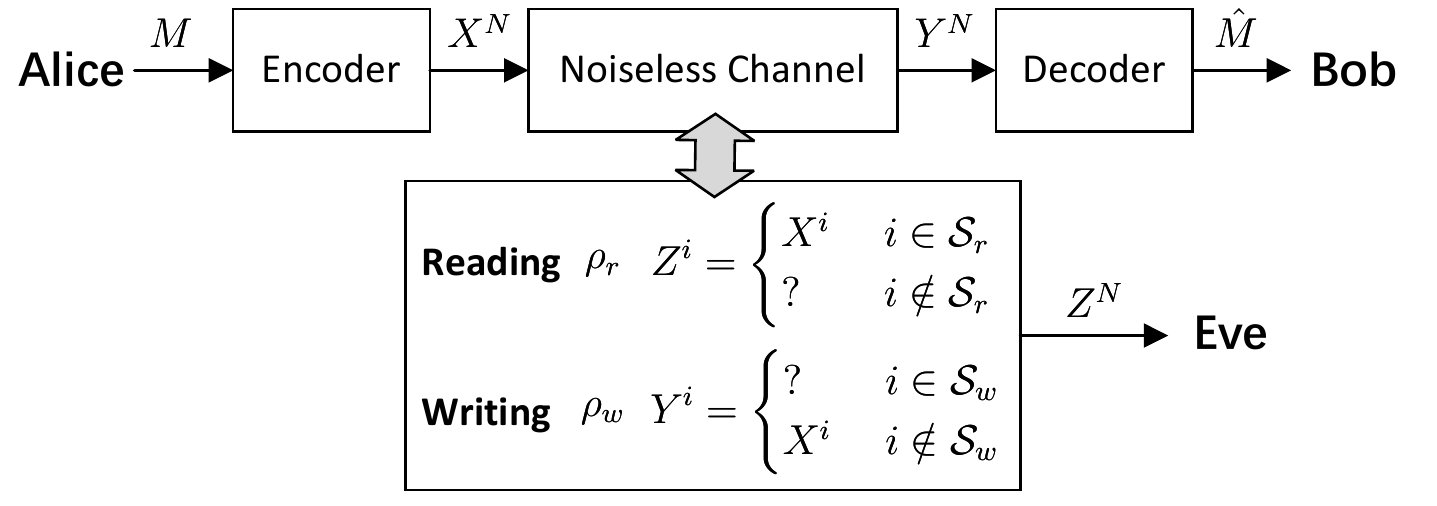}
\caption{The adversarial wiretap channel model.}
\label{fig_awt}
\end{figure}

The communication process over AWTC is illustrated in Fig.~\ref{fig_awt}. Legitimate user Alice want to send confidential message $M$ to legitimate user Bob with the existence of an active eavesdropper Eve. Alice encodes the message $M$ into channel input $X^N$ and transmits $X^N$ to Bob through a noiseless main channel. Eve arbitrarily reads $X^{\mathcal{S}_r}$ with fixed rate $\rho_r$ and obtains the corresponding $Z^N$. Eve also arbitrarily writes $X^{\mathcal{S}_w}$ with fixed rate $\rho_w$. Then Bob receives the modified channel output as $Y^N$ and decodes it into estimated confidential message $\hat{M}$.

\begin{definition}\label{def_criterion} For any $(2^{NR},N)$ secure codes of AWTC, the performances can be measured as follow.
\begin{itemize}
\item Reliability is measured by error probability $\mathrm{P_e}=\Pr(M\neq\hat{M})$. The reliability criterion is $\lim_{N\rightarrow\infty}\mathrm{P_e}=0$.
\item Security is measured by information leakage $\mathrm{L}=I(Z^N;M)$. The weak security criterion is $\lim_{N\rightarrow\infty}\frac{\mathrm{L}}{N}=0$; the strong security criterion is $\lim_{N\rightarrow\infty}\mathrm{L}=0$.
\end{itemize}
\end{definition}

The secrecy capacity of AWTC under reliability and security criterions has been characterized in \cite{Wang2015}.

\begin{theorem}\label{theo_cs}(Secrecy capacity\cite{Wang2015}) The perfect secrecy capacity of the AWTC with $(\rho_r,\rho_w)$ is
\begin{equation}
\mathrm{C_s}=1-\rho_r-\rho_w.
\end{equation}
\end{theorem}

In this paper, we intend to present a polar codes based solution for the problem of secure and reliable communication over AWTC model, and try to achieve the secrecy capacity of Theo.~\ref{theo_cs}.

\section{Polarization of Adversarial Wiretap Channels}\label{sec_polarization}

\subsection{Preliminaries on Channel Polarization}

\subsubsection{Polarization of Stationary Channels}

In Arikan's theory of channel polarization\cite{Arikan2009}, the channel polarization transformation $\mathbf{G}_N$ synthesizes from a kernel operation pair $(W,W)\mapsto(W^-,W^+)$ defined as follows. Consider two independent copies of channel $W:\{0,1\}\rightarrow\mathcal{Y}$, then we have the operation pair
\begin{equation}
\begin{split}
W^-=W\boxast W,\\
W^+=W\circledast W,
\end{split}
\end{equation}
specifically as
\begin{equation}
\begin{split}
&W^-(y^1,y^2|u^1)=\sum_{u^2\in \{0,1\}}\frac{1}{2}W(y^1|u^1\oplus u^2)W(y^2|u^2),\\
&W^+(y^1,y^2,u^1|u^2)=\frac{1}{2}W(y^1|u^1\oplus u^2)W(y^2|u^2).
\end{split}
\end{equation}

\begin{definition}
To measure the reliability of B-DMC $W$, Ar{\i}kan used the Bhattacharyya parameter $Z(W)$ defined as
\begin{equation}
Z(W)\triangleq\sum_{y\in\mathcal{Y}}\sqrt{W(y|0)W(y|1)}.
\end{equation}
Particularly for BEC, $I(W)=1-Z(W)$. And for channel sequence $W^{1:N}$, $\mathbb{Z}(W^{1:N})$ defines the sequence of $Z(W_1),Z(W_2),...,Z(W_N)$.
\end{definition}

Then for the kernel operation pair $(W,W)\mapsto(W^-,W^+)$, the Bhattacharyya parameter satisfies
\begin{equation}
\begin{split}
&Z(W^-)\leq 2Z(W)-Z(W)^2,\\
&Z(W^+)=Z(W)^2,\\
\end{split}
\end{equation}
with equality if $W$ is a BEC.

Let $W^N$ be $N$ independent copies of B-DMC $W$, $N=2^n$. Then the channel polarization transformation $\mathbf{G}_N$ is a recursive process of performing the kernel operation $(W,W)\mapsto(W^-,W^+)$ on $W^N$ as follows. For all $q\in[\![0,n-1]\!]$, $Q=2^q$, $i\in[\![1,Q]\!]$, with initial $W_1^{(i)}=W$, $Z(W_1^{(i)})=Z(W)$, have
\begin{equation}
(W_{Q}^{(i)},W_{Q}^{(i)})\mapsto(W_{2Q}^{(2i-1)},W_{2Q}^{(2i)}),
\label{eq_recu}
\end{equation}
where
\begin{equation}
\begin{split}
&W_{2Q}^{(2i-1)}=W_{Q}^{(i)}\boxast W_{Q}^{(i)},\\
&W_{2Q}^{(2i)}=W_{Q}^{(i)}\circledast W_{Q}^{(i)}.
\end{split}
\end{equation}
And satisfies
\begin{equation}\label{eq_rez_polar}
\begin{split}
&Z(W_{2Q}^{(2i-1)})\leq 2Z(W_{Q}^{(i)})-Z(W_{Q}^{(i)})^2,\\
&Z(W_{2Q}^{(2i)})=Z(W_{Q}^{(i)})^2,\\
\end{split}
\end{equation}
with equality if $W$ is a BEC.

By above recursive process, we can obtain the generated channels $W_N^{(1:N)}$ and the polarized Bhattacharyya parameters $Z(W_N^{(1:N)})$.

\subsubsection{Polarization of Non-stationary Channels}

%\begin{theorem}\label{theo_polarization} (Non-stationary polarization\cite[Theo.~2]{Zhao2020}) For any B-DMC $W^{1:N}$ with different transition probabilities, the generated channels ${W_N^{(i)}}$ form non-stationary channel transformation $\mathbf{G}_N$ are polarized in the sense that, for any fixed $\delta\in(0,1)$, as $N\rightarrow\infty$, the fraction of indices $i\in[\![1,N]\!]$  for which $I(W_N^{(i)})\in (1-\delta,1]$ goes to $\mathbb{A}[I(W^{1:N })]$ and the fraction for which $I(W_N^{(i)})\in [0,\delta)$ goes to $1-\mathbb{A}[I(W^{1:N })]$. Also can be write as
%\begin{equation}
%I_\infty=
%\begin{cases}
%1 &~\text{w.p.} ~\mathbb{A}[I(W^{1:N })]\\
%0 &~\text{w.p.}~1-\mathbb{A}[I(W^{1:N })],
%\end{cases}
%\end{equation}
%where $\mathbb{A}[I(W^{1:N })]$ is the average of the initial $I(W^{(i)})$ for all the $i\in[\![1,N]\!]$.
%\end{theorem}

The polarization theory of non-stationary channels has been studied in \cite{Zhao2020,Mahdavifar2018,Alsan2016}. These work proved that Ar{\i}kan's channel transformation $\mathbf{G}_N$ can also polarize non-stationary channels.

Let $W^1:\{0,1\}\rightarrow\mathcal{Y}^1$, $W^2:\{0,1\}\rightarrow\mathcal{Y}^2$ be two independent channels. Then for non-stationary channel, the kernel operation pair of channel polarization is $(W^1,W^2)\mapsto(\tilde{W}^-,\tilde{W}^+)$ that
\begin{equation}
\begin{split}
\tilde{W}^-=W^1\boxast W^2,\\
\tilde{W}^+=W^1\circledast W^2,
\end{split}
\end{equation}
specifically as
\begin{equation}
\begin{split}
&\tilde{W}^-(y^1,y^2|u^1)=\sum_{u^2\in \{0,1\}}\frac{1}{2}W^1(y^1|u^1\oplus u^2)W^2(y^2|u^2),\\
&\tilde{W}^+(y^1,y^2,u^1|u^2)=\frac{1}{2}W^1(y^1|u^1\oplus u^2)W^2(y^2|u^2).
\end{split}
\end{equation}
And the Bhattacharyya parameter satisfies \cite[Lem.~2]{Zhao2020}
\begin{equation}
\begin{split}
&Z(\tilde{W}^-)\leq Z(W^1)+Z(W^2)-Z(W^1)Z(W^2),\\
&Z(\tilde{W}^+)=Z(W^1)Z(W^2),\\
\end{split}
\end{equation}
with equality if $W^1$ and $W^2$ are BECs.

Let $W^{1:N}$ be a $N$-length non-stationary B-DMC sequence, $N=2^n$. Then the channel polarization transformation $\mathbf{G}_N$ on $W^{1:N}$ is a recursive process of performing the kernel operation $(W^1,W^2)\mapsto(\tilde{W}^-,\tilde{W}^+)$ as follows. For for all $q\in[\![0,n-1]\!]$, $Q=2^q$, $k\in[\![0,\frac{N}{2Q}-1]\!]$, $i\in[\![2kQ+1,2kQ+Q]\!]$, with initial $W_1^{(i)}=W^i$, $Z(W_1^{(i)})=Z(W^i)$, have
\begin{equation}
(W_{Q}^{(i)},W_{Q}^{(i+Q)})\mapsto(W_{2Q}^{(2i-2kQ-1)},W_{2Q}^{(2i-2kQ)}),
\end{equation}
where
\begin{equation}
\begin{split}
&W_{2Q}^{(2i-2kQ-1)}=W_{Q}^{(i)}\boxast W_{Q}^{(i+Q)},\\
&W_{2Q}^{(2i-2kQ)}=W_{Q}^{(i)}\circledast W_{Q}^{(i+Q)}.
\end{split}
\end{equation}
And satisfies \begin{equation}
\begin{split}
&Z(W_{2Q}^{(2i-2kQ-1)})\leq Z(W_{Q}^{(i)})+Z(W_{Q}^{(i+Q)})-Z(W_{Q}^{(i)})Z(W_{Q}^{(i+Q)}),\\
&Z(W_{2Q}^{(2i-2kQ)})=Z(W_{Q}^{(i)})Z(W_{Q}^{(i+Q)}),\\
\end{split}
\end{equation}
with equality if all $W^{1:N}$ are BECs.

By above recursive process, we can obtain the generated channels $W_N^{(1:N)}$ and the corresponding polarized Bhattacharyya parameters sequence $\mathbb{Z}(W_N^{(1:N)})$.

\textbf{Remark.} Note that for the recursive process of non-stationary channels polarization, by assuming the initial channels $W^{1:N}$ are identical, the whole process will be the same as the polarization process of stationary channels.

\subsection{BEC Based Equivalent AWTC Model}

To study the polarization of AWTC model, as the first step, we have to the channel based equivalent models for the direct reading and writing actions in AWTC.

Consider a binary input AWTC model in Def.~\ref{def_awt} that $\mathcal{X}=\{0,1\}$, $\mathcal{Y}=\{0,1,?\}$ and $\mathcal{Z}=\{0,1,?\}$. To establish the equivalent models, we use two kinds of bit channels $V_{\epsilon_1}$ and $V_{\epsilon_0}$ which are defined as follows.
\begin{definition}
Define $V_{\epsilon_1}:\{0,1\}\rightarrow\{0,1,?\}$ the full-noise symmetric binary erase channel (BEC) with erasure probability $\epsilon=1$ that $I(V_{\epsilon_1})=0$. Define $V_{\epsilon_0}:\{0,1\}\rightarrow\{0,1,?\}$ the noiseless symmetric BEC with erasure probability $\epsilon=0$ that $I(V_{\epsilon_0})=1$.
\end{definition}

Then for the AWTC model, we can observe that
\begin{itemize}
\item For adversary Eve, the bits that she directly reads are equivalent to being transmitted to her through the noiseless bit BEC $V_{\epsilon_0}$; and the bits that she dose not read are equivalent to being transmitted to her through the full-noise bit BEC $V_{\epsilon_1}$.
\item For legitimate receiver Bob, the bits that Eve writes are equivalent to being transmitted to him through the full-noise bit BEC $V_{\epsilon_1}$ ; and the bits that Eve dose not write are equivalent to being transmitted to him through the noiseless bit BEC $V_{\epsilon_0}$ .
\end{itemize}

Therefore, we have the BEC based equivalent AWTC models as follows.

\begin{definition}\label{def_eqvawtc} Let The BEC based equivalent AWTC model is defined as $(W_w^{1:N},W_r^{1:N}):\mathcal{X}^N\rightarrow \mathcal{Y}^N, \mathcal{Z}^N$ that
\begin{itemize}

\item $W_w^{1:N}:\mathcal{X}^N\rightarrow \mathcal{Y}^N$ is the writing-equivalent main channel block. For arbitrarily chosen $\mathcal{S}_w\subseteq[\![1,N]\!]$ with fixed fraction $\rho_w$, have
\begin{equation}
W_w^i=
\begin{cases}
V_{\epsilon_1}&~\text{if}~i\in \mathcal{S}_w\\
V_{\epsilon_0}&~\text{if}~i\in \mathcal{S}_w^c.
\end{cases}
\end{equation}
\item $W_r^{1:N}:\mathcal{X}^N\rightarrow \mathcal{Z}^N$ is the reading-equivalent wiretap channel block. For arbitrarily chosen $\mathcal{S}_r\subseteq[\![1,N]\!]$ with fixed fraction $\rho_r$, have
\begin{equation}
W_r^i=
\begin{cases}
V_{\epsilon_0}&~\text{if}~i\in \mathcal{S}_r\\
V_{\epsilon_1}&~\text{if}~i\in \mathcal{S}_r^c.
\end{cases}
\end{equation}
\end{itemize}
\end{definition}

Then from Def.~\ref{def_eqvawtc}, for both $W_w^{1:N}$ and $W_r^{1:N}$, we can further conclude an unified expression as the $\rho$ equivalent channel block which is defined as follows.

\begin{definition}\label{def_rec} The $\rho$ equivalent channel block is defined as $W_\rho^{1:N}:\mathcal{X}^N\rightarrow \mathcal{Y}^N$ that for arbitrarily chosen $\mathcal{S}_\rho\subseteq[\![1,N]\!]$ with fixed fraction $\rho$,
\begin{equation}
W_\rho^i=
\begin{cases}
V_{\epsilon_1}&~\text{if}~i\in \mathcal{S}_\rho\\
V_{\epsilon_0}&~\text{if}~i\in \mathcal{S}_\rho^c.
\end{cases}
\end{equation}
The average erase probability of the channel block is $\rho$.
\end{definition}

\textbf{Remark.} $W_\rho^{1:N}$ can be used to describe both writing-equivalent channel block $W_w^{1:N}$ and reading-equivalent channel block $W_r^{1:N}$. If we let $\rho=\rho_w$, then  $W_\rho^{1:N}=W_{\rho_w}^{1:N}$ which is same as $W_w^{1:N}$. If we let $\rho=1-\rho_r$, then $W_\rho^{1:N}=W_{1-\rho_r}^{1:N}$ which is same as $W_r^{1:N}$. Therefore, to study the polarization of AWTC model, we only have to analyze the polarization of the $\rho$ equivalent channel block under operation $\mathbf{G}_N$.

\subsection{Polarization of the $\rho$ Equivalent Channel Block}

As defined in Def.~\ref{def_rec}, the $\rho$ equivalent channel block $W_\rho^{1:N}$ is an $N$-length \emph{non-stationary channel sequence} consisted of full noise BEC $V_{\epsilon_1}$ with fraction $\rho$ and noiseless BEC $V_{\epsilon_0}$ with fraction $1-\rho$. Thus if we apply the channel operation $\mathbf{G}_N$ on $\rho$ equivalent channel block, the channel transformation process follows the kernel operation $(W^1,W^2)\mapsto(\tilde{W}^-,\tilde{W}^+)$ of non-stationary channel polarization. The recursive polarization process from $W_\rho^{1:N}$ to $W_{\rho, N}^{(1:N)}$ is as follows.

For for all $q\in[\![0,n-1]\!]$, $Q=2^q$, $k\in[\![0,\frac{N}{2Q}-1]\!]$, $i\in[\![2kQ+1,2kQ+Q]\!]$, with initial $W_{\rho,1}^{(i)}=W_{\rho}^i$, $Z(W_{\rho,1}^{(i)})=Z(W_\rho^i)$, have
\begin{equation}\label{eq_re_rho}
(W_{\rho,Q}^{(i)},W_{\rho,Q}^{(i+Q)})\mapsto(W_{\rho,2Q}^{(2i-2kQ-1)},W_{\rho,2Q}^{(2i-2kQ)}),
\end{equation}
where
\begin{equation}
\begin{split}
&W_{\rho,2Q}^{(2i-2kQ-1)}=W_{\rho,Q}^{(i)}\boxast W_{\rho,Q}^{(i+Q)},\\
&W_{\rho,2Q}^{(2i-2kQ)}=W_{\rho,Q}^{(i)}\circledast W_{\rho,Q}^{(i+Q)}.
\end{split}
\end{equation}
Thus we have the iteration of the Bhattacharyya parameter as
\begin{equation}\label{eq_rez_rho}
\begin{split}
&Z(W_{\rho,2Q}^{(2i-2kQ-1)})= Z(W_{\rho,Q}^{(i)})+Z(W_{\rho,Q}^{(i+Q)})-Z(W_{\rho,Q}^{(i)})Z(W_{\rho,Q}^{(i+Q)}),\\
&Z(W_{\rho,2Q}^{(2i-2kQ)})=Z(W_{\rho,Q}^{(i)})Z(W_{\rho,Q}^{(i+Q)}).\\
\end{split}
\end{equation}

Note that for the initial $\mathbb{Z}(W_{\rho,1}^{(1:N)})$, $Z(W_{\rho,1}^{(j)})$ with any $j\in[\![1,N]\!]$ is either $V_{\epsilon_1}$ or $V_{\epsilon_0}$. So for $Z(W_{\rho,1}^{(j)})$, there are only two possible values, $Z(W_{\rho,1}^{(j)})=1$ with probability $\rho$ and $Z(W_{\rho,1}^{(j)})=0$ with probability $1-\rho$. To measure this probability of Bhattacharyya parameter, we use the $P(W)$ defined in the following.

\begin{definition}
For any channel $W$, $P(W)$ is defined as the probability $\Pr\{Z(W)=1\}$. Particularly, for channel sequence $W^{1:N}$, $\mathbb{P}(W^{1:N})$ is defined as the sequence of $P(W^1),P(W^2),...,P(W^N)$.
\end{definition}

Therefore, initially for all $j\in[\![1,N]\!]$, we have $P(W_{\rho,1}^{(j)})=\rho$. Then for the iteration of \eqref{eq_rez_rho} that $[Z(W_{\rho,Q}^{(i)}),Z(W_{\rho,Q}^{(i+Q)})]\mapsto[Z(W_{\rho,2Q}^{(2i-2kQ-1)}),Z(W_{\rho,2Q}^{(2i-2kQ)})]$, there are four possible cases, namely as $(1,1)\mapsto(1,1)$, $(1,0)\mapsto(1,0)$, $(0,1)\mapsto(1,0)$ and $(0,0)\mapsto(0,0)$. Thus for all the channels in the iteration, considering the probabilities $P(W_{\rho,Q}^{(i)})$,$P(W_{\rho,Q}^{(i+Q)})$,$P(W_{\rho,2Q}^{(2i-2kQ-1)})$ and $P(W_{\rho,2Q}^{(2i-2kQ)})$, we have
\begin{equation}\label{eq_rep_rho}
\begin{split}
P(W_{\rho,2Q}^{(2i-2kQ-1)})&=1-[1-P(W_{\rho,Q}^{(i)})][1-P(W_{\rho,Q}^{(i+Q)})]\\
&=P(W_{\rho,Q}^{(i)})+P(W_{\rho,Q}^{(i+Q)})-P(W_{\rho,Q}^{(i)})P(W_{\rho,Q}^{(i+Q)}),\\
P(W_{\rho,2Q}^{(2i-2kQ)})&=P(W_{\rho,Q}^{(i)})P(W_{\rho,Q}^{(i+Q)}).\\
\end{split}
\end{equation}

So we can observe that $[P(W_{\rho,Q}^{(i)}),P(W_{\rho,Q}^{(i+Q)})]\mapsto[P(W_{\rho,2Q}^{(2i-2kQ-1)}),P(W_{\rho,2Q}^{(2i-2kQ)})]$ in \eqref{eq_rep_rho} has a same iteration process as $[Z(W_{\rho,Q}^{(i)}),Z(W_{\rho,Q}^{(i+Q)})]\mapsto[Z(W_{\rho,2Q}^{(2i-2kQ-1)}),Z(W_{\rho,2Q}^{(2i-2kQ)})]$ in \eqref{eq_rez_rho}.

Since initially all $j\in[\![1,N]\!]$, $P(W_{\rho,1}^{(j)})=\rho$, \eqref{eq_rep_rho} can be further simplified as follows. Let $N=2^n$ and $W_{\rho,N}^{(1:N)}$ is the channel block generated from $W_\rho^{1:N}$ by non-stationary channel operation $\mathbf{G}_N$, then the corresponding $\mathbb{P}(W_{\rho,N}^{(1:N)})$ can be calculated by following iterations. For all $q\in[\![0,n-1]\!]$, $Q=2^q$, $i\in[\![1,Q]\!]$, with initial $P(W_{\rho,1}^{(i)})=\rho$, have
\begin{equation}\label{eq_s_rep_rho}
\begin{split}
&P(W_{\rho,2Q}^{(2i-1)})= 2P(W_{\rho,Q}^{(i)})-P(W_{\rho,Q}^{(i)})^2,\\
&P(W_{\rho,2Q}^{(2i)})=P(W_{\rho,Q}^{(i)})^2.\\
\end{split}
\end{equation}

Next, to analyze above process from $\mathbb{P}(W_{\rho,1}^{(1:N)})$ to $\mathbb{P}(W_{\rho,N}^{(1:N)})$, lets consider a stationary BEC block $V_{\epsilon_\rho}^N$ defined in Def.~\ref{def_rbec} and its polarization process as an analogy.
\begin{definition}\label{def_rbec}
Define $V_{\epsilon_\rho}:\{0,1\}\rightarrow\{0,1,?\}$ the BEC with erasure probability $\epsilon=\rho$, have $Z(V_{\epsilon_\rho})=\rho$. Let $V_{\epsilon_\rho}^N$ be the $N$-length stationary channel block of $V_{\epsilon_\rho}$.
\end{definition}

For BEC block $V_{\epsilon_\rho}^N$, the stationary polarization process from $\mathbb{Z}(V_{\epsilon_\rho,1}^{(1:N)})$ to $\mathbb{Z}(V_{\epsilon_\rho,N}^{(1:N)})$ is as follows.
For all $q\in[\![0,n-1]\!]$, $Q=2^q$, $i\in[\![1,Q]\!]$, with initial $Z(V_{\epsilon_\rho,1}^{(i)})=\rho$, have
\begin{equation}\label{eq_rez_v}
\begin{split}
&Z(V_{\epsilon_\rho,2Q}^{(2i-1)})= 2Z(V_{\epsilon_\rho,Q}^{(i)})-Z(V_{\epsilon_\rho,Q}^{(i)})^2,\\
&Z(V_{\epsilon_\rho,2Q}^{(2i)})=Z(V_{\epsilon_\rho,Q}^{(i)})^2.\\
\end{split}
\end{equation}

\emph{Comparing \eqref{eq_s_rep_rho} with \eqref{eq_rez_v}, the iteration process from $\mathbb{P}(W_{\rho,1}^{(1:N)})$ to $\mathbb{P}(W_{\rho,N}^{(1:N)})$ is exactly the same as the polarization process from $\mathbb{Z}(V_{\epsilon_\rho,1}^{(1:N)})$ to $\mathbb{Z}(V_{\epsilon_\rho,N}^{(1:N)})$. So by $P(W_{\rho,1}^{(i)})=Z(V_{\epsilon_\rho,1}^{(i)})=\rho$, we have $\mathbb{P}(W_{\rho,N}^{(1:N)})=\mathbb{Z}(V_{\epsilon_\rho,N}^{(1:N)})$, which concludes that $\mathbb{P}(W_{\rho,N}^{(1:N)})$  has the same polarization results as $\mathbb{Z}(V_{\epsilon_\rho,N}^{(1:N)})$ by the channel polarization operation $\mathbf{G}_N$.}

\begin{proposition}\label{prop_polarization}
For the $\rho$ equivalent channel block $W_\rho^{1:N}$ with a fixed $\rho$, the channels $W_{\rho,N}^{(1:N)}$ polarize in the sense that, for any fixed $\delta_N=2^{-N^\beta}$, $\beta\in(0,\frac{1}{2})$, as $N=2^n$ goes to infinity, the fraction of indices $i\in[\![1,N]\!]$ for which $P(W_{\rho,N}^{(i)})\in[0,\delta_N)$ goes to $1-\rho$ and the fraction of indices $i\in[\![1,N]\!]$ for which $P(W_{\rho,N}^{(i)})\in(1-\delta_N,1]$ goes to $\rho$.
\begin{IEEEproof}
Note that for BEC block $V_{\epsilon_\rho}^N$, we have the polarized subsets of index $[\![1,N]\!]$ as
\begin{equation}
\begin{split}
&\mathcal{H}_{\epsilon_\rho}=\{ i\in [\![1,N]\!]: Z(V_{\epsilon_\rho,N}^{(i)})\geq1-\delta_{N}\},\\
&\mathcal{L}_{\epsilon_\rho}=\{ i\in [\![1,N]\!]: Z(V_{\epsilon_\rho,N}^{(i)})\leq\delta_{N}\}.
\end{split}
\end{equation}
where $\mathcal{H}_{\epsilon_\rho}$ is the index subset of almost full-noise channels, and $\mathcal{L}_{\epsilon_\rho}$ is the index subset of almost noiseless channels.

Since $\mathbb{P}(W_{\rho,N}^{(1:N)})=\mathbb{Z}(V_{\epsilon_\rho,N}^{(1:N)})$, for the polarization of the $\rho$ equivalent channel block $W_\rho^{1:N}$, we can have
\begin{equation}
\begin{split}
&\mathcal{H}_\rho=\{ i\in [\![1,N]\!]: P(W_{\rho,N}^{(i)})\geq1-\delta_{N}\},\\
&\mathcal{L}_\rho=\{ i\in [\![1,N]\!]: P(W_{\rho,N}^{(i)})\leq\delta_{N}\}.
\end{split}
\end{equation}
where $\mathcal{H}_\rho$ is the subset in which  $\Pr\{Z(W_{\rho,N}^{(i)})=1\}$ is almost $1$, and $\mathcal{L}_\rho$ is the subset in which $\Pr\{Z(W_{\rho,N}^{(i)})=1\}$ is almost $0$ (same as $\Pr\{Z(W_{\rho,N}^{(i)})=0\}$ is almost $1$).

Also, we can observe that $\mathcal{H}_\rho=\mathcal{H}_{\epsilon_\rho}$ and $\mathcal{L}_\rho=\mathcal{L}_{\epsilon_\rho}$. Thus we have
\begin{equation}\label{eq_rate}
\begin{split}
\lim_{N\rightarrow\infty}\frac{1}{N}|\mathcal{H}_\rho|&=\lim_{N\rightarrow\infty}\frac{1}{N}|\mathcal{H}_{\epsilon_\rho}|\\
&=1-I(V_{\epsilon_\rho})\\
&=\rho,\\
\lim_{N\rightarrow\infty}\frac{1}{N}|\mathcal{L}_\rho|&=\lim_{N\rightarrow\infty}\frac{1}{N}|\mathcal{L}_{\epsilon_\rho}|\\
&=I(V_{\epsilon_\rho})\\
&=1-\rho.\\
\end{split}
\end{equation}
\end{IEEEproof}
\end{proposition}

%\subsection{Decoding of the $\rho$ Equivalent Channel Block}
%
%Recall that for polar codes, Ar{\i}kan basic
%
%\begin{itemize}
%\item $y^i=?$:
%\begin{equation}
%\begin{split}
%L_1^{(1)}(?)&=\frac{\rho V_{\epsilon_1}(?|0)+(1-\rho) V_{\epsilon_0}(?|0)}{\rho V_{\epsilon_1}(?|1)+(1-\rho) V_{\epsilon_0}(?|1)}\\
%&=\frac{\rho \cdot 1+(1-\rho) \cdot 0}{\rho \cdot 1+(1-\rho) \cdot 0}=1
%\end{split}
%\end{equation}
%
%\item $y^i=0$:
%\begin{equation}
%\begin{split}
%L_1^{(1)}(0)&=\frac{\rho V_{\epsilon_1}(0|0)+(1-\rho) V_{\epsilon_0}(0|0)}{\rho V_{\epsilon_1}(0|1)+(1-\rho) V_{\epsilon_0}(0|1)}\\
%&=\frac{\rho \cdot 0+(1-\rho) \cdot 1}{\rho \cdot 0+(1-\rho) \cdot 0}\approx \frac{1}{\gamma}
%\end{split}
%\end{equation}
%where $\gamma$ is a sufficiently small positive number.
%
%\item $y^i=1$:
%\begin{equation}
%\begin{split}
%L_1^{(1)}(1)&=\frac{\rho V_{\epsilon_1}(1|0)+(1-\rho) V_{\epsilon_0}(1|0)}{\rho V_{\epsilon_1}(1|1)+(1-\rho) V_{\epsilon_0}(1|1)}\\
%&=\frac{\rho \cdot 0+(1-\rho) \cdot 0}{\rho \cdot 0+(1-\rho) \cdot 1}=0
%\end{split}
%\end{equation}
%\end{itemize}

\section{Secure Polar Codes}\label{sec_code}

In this section, we present the secure polar codes for the AWTC model based on the equivalent AWTC model and the polarization result of the $\rho$ equivalent channel block $W_\rho^{1:N}$.

\subsection{Secure Polar Coding Scheme}

In the AWTC model, the fraction pair $(\rho_w,\rho_r)$ of adversarial writing and reading is publicly known and fixed. For adversarial writing, the equivalent channel block is $W_w^{1:N}$ which also is $W_{\rho_w}^{1:N}$ in the form of $\rho$ equivalent channel block. For adversarial reading, the equivalent channel block is $W_r^{1:N}$ which is also $W_{1-\rho_r}^{1:N}$ in the form of $\rho$ equivalent channel block. Then by non-stationary channel polarization operation $\mathbf{G}_N$, both $W_w^{1:N}$ and $W_r^{1:N}$ can be polarized as follows.

For the writing equivalent channel block $W_w^{1:N}$ (also as $W_{\rho_w}^{1:N}$), with any fixed $\delta_N=2^{-N^\beta}$, $\beta\in(0,\frac{1}{2})$, the index $[\![1,N]\!]$ can be polarized into $\mathcal{H}_w$ and $\mathcal{L}_w$ that
\begin{equation}\label{eq_w_polar}
\begin{split}
&\mathcal{H}_w=\{ i\in [\![1,N]\!]: P(W_{\rho_w,N}^{(i)})\geq1-\delta_{N}\},\\
&\mathcal{L}_w=\{ i\in [\![1,N]\!]: P(W_{\rho_w,N}^{(i)})\leq\delta_{N}\}.\\
\end{split}
\end{equation}
For legitimate user Bob, every channel in $\mathcal{H}_w$ is full noise with probability almost $1$; and every channel in $\mathcal{L}_w$ is noiseless with probability almost $1$.

For the reading equivalent channel block $W_r^{1:N}$ (also as $W_{1-\rho_r}^{1:N}$), with any fixed $\delta_N=2^{-N^\beta}$, $\beta\in(0,\frac{1}{2})$, the index $[\![1,N]\!]$ can be polarized into $\mathcal{H}_r$ and $\mathcal{L}_r$ that
\begin{equation}\label{eq_r_polar}
\begin{split}
&\mathcal{H}_r=\{ i\in [\![1,N]\!]: P(W_{1-\rho_r,N}^{(i)})\geq1-\delta_{N}\},\\
&\mathcal{L}_r=\{ i\in [\![1,N]\!]: P(W_{1-\rho_r,N}^{(i)})\leq\delta_{N}\}.
\end{split}
\end{equation}
For adversary Eve, every channel in $\mathcal{H}_r$ is full noise with probability almost $1$; and every channel in $\mathcal{L}_r$ is noiseless with probability almost $1$.

Based on the polarization results of $W_w^{1:N}$ and $W_r^{1:N}$, we can apply the multi-block chaining structure\cite{Vard2013strong} to construct a strong security polar coding scheme. Firstly, divided the index $[\![1,N]\!]$ into following four subsets:
\begin{equation}\label{eq_dv1}
\begin{split}
\mathcal{I}=\mathcal{L}_w \cap \mathcal{H}_r,\\
\mathcal{R}=\mathcal{L}_w \cap \mathcal{H}_r^c,\\
\mathcal{F}=\mathcal{L}_w^c \cap \mathcal{H}_r,\\
\mathcal{B}=\mathcal{L}_w^c \cap \mathcal{H}_r^c.
\end{split}
\end{equation}

Then form subset $\mathcal{I}$, separate a subset $\mathcal{E}$ that satisfies $|\mathcal{E}|=|\mathcal{B}|$. Finally we obtain the subsets group $(\mathcal{I}\setminus\mathcal{E},\mathcal{E},\mathcal{R},\mathcal{F},\mathcal{B})$ that specifies as follows:
\begin{itemize}
  \item $\mathcal{I}\setminus\mathcal{E}$ is for information bits, because the corresponding channels are secure and reliable with probability close to $1$.
  \item $\mathcal{E}$ is for uniformly distributed random bits for the subset $\mathcal{B}$ of next block, because the corresponding channels are secure and reliable with probability close to $1$.
  \item $\mathcal{R}$ is for uniformly distributed random bits, because the corresponding channels are reliable but insecure with probability close to $1$.
  \item $\mathcal{F}$ is for publicly known frozen bits to guarantee the reliability, because the corresponding channels are secure but unreliable with probability close to $1$.
  \item $\mathcal{B}$ are neither secure nor reliable with probability close to $1$. In the first block, it is for random bits pre-shared by legitimate parties. In the rest blocks, it is for the bit of $\mathcal{E}$ in the previous block. Therefore both reliability and security can be guaranteed.
\end{itemize}

Now we present the secure polar coding scheme. Consider the multi-block case of AWTC with block length $N=2^n$, block number $T$ and fixed fraction pair $(\rho_w,\rho_r)$. For block $t\in[\![1,T]\!]$,
\begin{itemize}
\item \emph{Encoding:}
    \begin{itemize}
      \item $u^{\mathcal{I}\setminus\mathcal{E}}$ are assigned with information bits;
      \item $u^{\mathcal{E}\cup\mathcal{R}}$ are assigned with uniformly distributed random bits;
      \item $u^{\mathcal{F}}$ are assigned with publicly known frozen bits;
      \item if $t=1$, $u^{\mathcal{B}}$ are assigned with pre-shared random bits;
      \item if $t\geq2$, $u^{\mathcal{B}}$ are assigned with the bits of $u^{\mathcal{E}}$ of block $t-1$;
      \item encode the $u^N$ into channel inputs $x^N$ by $x^N=u^N\mathbf{G}_N$.
    \end{itemize}
\item \emph{Transmission:}
    \begin{itemize}
      \item Alice transmits $x^N$ to Bob through a noiseless communication channel;
      \item Eve arbitrarily chooses $\mathcal{S}_w$ with fraction $\rho_w$ and writes $x^{\mathcal{S}_w}$ into $``?"$;
      \item Bob receives the modified channel outputs as $y^N$;
      \item Eve arbitrarily chooses $\mathcal{S}_r$ with fraction $\rho_r$ and reads $x^{\mathcal{S}_r}$ to obtain $z^N$.
    \end{itemize}
\item \emph{Decoding:} Bob decodes $y^N$ into $\hat{u}^N$ by successive cancelation decoding\cite{Arikan2009} as follow:
    \begin{itemize}
      \item if $i\in\mathcal{I}\cup\mathcal{R}$,
      \begin{equation}
      \hat{u}^i=\arg \max\limits_{u\in\{0,1\}} W_{\rho_w, N}^{(i)}(u|\hat{u}^{1:i-1},y^{1:N}),
      \end{equation}
      with initial likelihood ratio
      \begin{equation}
      L_1^{(i)}(y^i)=\frac{\rho_w V_{\epsilon_1}(y^i|0)+(1-\rho_w) V_{\epsilon_0}(y^i|0)}{\rho_w V_{\epsilon_1}(y^i|1)+(1-\rho_w) V_{\epsilon_0}(y^i|1)};
      \end{equation}
      \item if $i\in\mathcal{F}$, $\hat{u}^i$ is decoded as publicly known frozen bit;
      \item if $i\in\mathcal{B}$, in case of $t=1$, $\hat{u}^i$ is decoded as pre-shared random bit, in case of $t\geq2$, $\hat{u}^i$ is decoded as corresponding bit of $\hat{u}^{\mathcal{E}}$ of block $t-1$.
    \end{itemize}
\end{itemize}

\subsection{Performance Analysis}\label{sec_performance}

Now we analyze the performance of proposed secure polar coding scheme by reliability, security and achievable secrecy rate.

\begin{lemma}\label{lem_drr}(Decoding error rate of polar codes\cite[Prop.~3]{Mahdavifar2011}) For any B-DMC channel block $W^N$, let $\mathcal{A}$ be an arbitrary subset of index $[\![1,N]\!]$ and used as the information set for polar codes. Then the corresponding block error rate of SC decoding satisfies
\begin{equation}
  \mathrm{P_e}\leq\sum_{i\in \mathcal{A}}Z(W_N^{(i)}).
\end{equation}
\end{lemma}

%\begin{theorem}\label{theo_rip} (Rate of non-stationary polarization\cite[Theo.~2]{Zhao2020})For any B-DMC $W^{1:N}$ with $I(W^i)>0$, and any fixed $R<\mathbb{A}[I(W^{1:N})]$ and constant $\beta\leq\frac{1}{2}$, there exists index set $\mathcal{A}\subset[\![1,N]\!]$, $|\mathcal{A}|\leq NR$ that
%\begin{equation}
%\sum_{i\in\mathcal{A}}Z(W_N^{(i)})=o(2^{-N^\beta})
%\end{equation}
%and
%\begin{equation}
%\mathrm{P_e}(N,R)=o(2^{-N^\beta}).
%\end{equation}
%\end{theorem}

\begin{proposition}\label{prop_reliability}With with $N\rightarrow\infty$, reliability can be achieved by the proposed secure polar coding scheme over the AWTC model.
\begin{IEEEproof}
Since the frozen bits in $\mathcal{F}$ are publicly known and Bob known the pre-shared bits for the first $\mathcal{B}$, then according to the multi-block chaining structure, the decoding error rate of entire $T$ blocks are determined by the SC decoding of $T$ blocks' $\mathcal{I}\cup\mathcal{R}$ and $T-1$ blocks' $\mathcal{E}$. Thus for the decoding error of $T$ blocks, by Lem.~\ref{lem_drr}, have
\begin{equation}\label{eq_ber}
\mathrm{P_e}(T)\leq T\sum_{i\in \mathcal{I}\cup\mathcal{R}}Z(W_{\rho_w, N}^{(i)})+(T-1)\sum_{i\in \mathcal{E}}Z(W_{\rho_w, N}^{(i)}).\\
\end{equation}
Consider probabilities
\begin{equation}
\begin{split}
\Pr\{\sum_{i\in\mathcal{I}\cup\mathcal{R}}Z(W_{\rho_w,N}^{(i)})=0\}&=\prod_{i\in\mathcal{I}\cup\mathcal{R}}[1-P(W_{\rho_w,N}^{(i)})]\\
&\geq(1-\delta_N)^{|\mathcal{I}\cup\mathcal{R}|},\\
\Pr\{\sum_{i\in \mathcal{E}}Z(W_{\rho_w, N}^{(i)})=0\}&=\prod_{i\in \mathcal{E}}[1-P(W_{\rho_w, N}^{(i)})]\\
&\geq(1-\delta_N)^{|\mathcal{E}|},
\end{split}
\end{equation}
since
\begin{equation}\label{eq_lim_deta}
\begin{split}
\lim_{N\rightarrow\infty}(1-\delta_N)^N&=\lim_{N\rightarrow\infty}(1-2^{-N^\beta})^N\\
&=\lim_{N\rightarrow\infty}\left[(1-\frac{1}{2^{N^\beta}})^{-2^{N^\beta}}\right]^{\frac{N}{-2^{N^\beta}}}\\
&=e^{\lim_{N\rightarrow\infty}\frac{N}{-2^{N^\beta}}}\\
&=e^0=1,
\end{split}
\end{equation}
we have
\begin{equation}
\begin{split}
&\lim_{N\rightarrow\infty}\Pr\{\sum_{i\in\mathcal{I}\cup\mathcal{R}}Z(W_{\rho_w,N}^{(i)})=0\}=1,\\
&\lim_{N\rightarrow\infty}\Pr\{\sum_{i\in\mathcal{E}}Z(W_{\rho_w,N}^{(i)})=0\}=1.\\
\end{split}
\end{equation}
Thus we have $\lim_{N\rightarrow\infty}\Pr\{\mathrm{P_e}(T)=0\}=1$, which proves that the reliability can be achieved.
\end{IEEEproof}
\end{proposition}

\begin{proposition}\label{prop_security}With $N\rightarrow\infty$, strong security can be achieved by the proposed secure polar coding scheme over the AWTC model.
\begin{IEEEproof}
For block $t$, let $\mathrm{M}^t=U^{\mathcal{I}\setminus\mathcal{E}}$, $\mathrm{Z}^t=Z^N$, $\mathrm{E}^t=U^{\mathcal{E}}$ and $\mathrm{F}^t=U^{\mathcal{F}}$. Then the information leakage for entire $T$ blocks is $\mathrm{L}(T)=I(\mathrm{M}^{1:T};\mathrm{Z}^{1:T})$.

For multi-block chaining structure, as deduced in \cite[Section IV-B]{Vard2013strong}, with publicly known frozen bits, have
\begin{equation}
\mathrm{L}(T)\leq\sum_{t=1}^T I(\mathrm{M}^t,\mathrm{E}^t,\mathrm{F}^t;\mathrm{Z}^t)+I(\mathrm{E}^0;\mathrm{Z}^0),
\end{equation}
where $I(\mathrm{E}^0;\mathrm{Z}^0)$ refers to the information leakage of the pre-shared bits before transmission which should be $0$.

Let $\mathrm{a}^1<\mathrm{a}^2<...<\mathrm{a}^{|\mathcal{A}|}$ be the correspondent indices of the elements $U^\mathcal{A}$ for any subset $\mathcal{A}$, such that $U^\mathcal{A}\triangleq U^{\mathrm{a}^1:\mathrm{a}^{|\mathcal{A}|}}=U^{\mathrm{a}^1},...,U^{\mathrm{a}^{|\mathcal{A}|}}$. Since subsets $\mathcal{I}_\epsilon\cup\mathcal{F}_\epsilon$ and $\mathcal{R}_\epsilon$ match the construction of induced channel \cite[Lem.~15]{Mahdavifar2011}, we have
\begin{equation}\label{eq_laekage}
\begin{split}
I(\mathrm{M}^t,\mathrm{E}^t,\mathrm{F}^t;\mathrm{Z}^t)
=&I(U^{\mathcal{I}\cup\mathcal{F}};Z^N)\\
=&\sum_{i=1}^{|\mathcal{I}\cup\mathcal{F}|}I(U^{\mathrm{a}^i};Z^N|U^{\mathrm{a}^1:\mathrm{a}^{i-1}})\\
\overset{(a)}{=}& \sum_{i=1}^{|\mathcal{I}\cup\mathcal{F}|}I(U^{\mathrm{a}^i};U^{\mathrm{a}^1:\mathrm{a}^{i-1}},Z^N)\\
\leq&\sum_{i=1}^{|\mathcal{I}\cup\mathcal{F}|}I(U^{\mathrm{a}^i};U^{1:\mathrm{a}^{i}-1},Z^N)\\
=& \sum_{j\in\mathcal{I}\cup\mathcal{F}}[1- Z(W_{1-\rho_r,N}^{(j)})]\\
\end{split}
\end{equation}
where $(a)$ is because $U^{\mathrm{a}^{i}}$ are independent from each other. Consider probability
\begin{equation}
\begin{split}
\Pr\{\sum_{j\in\mathcal{I}\cup\mathcal{F}}[1- Z(W_{1-\rho_r,N}^{(j)})]=0\}&=\prod_{j\in\mathcal{I}\cup\mathcal{F}}P(W_{1-\rho_r,N}^{(j)})\\
&\geq(1-\delta_N)^{|\mathcal{I}\cup\mathcal{F}|},
\end{split}
\end{equation}
by \eqref{eq_lim_deta}, we have
\begin{equation}
\lim_{N\rightarrow\infty}\Pr\{\sum_{j\in\mathcal{I}\cup\mathcal{F}}[1- Z(W_{1-\rho_r,N}^{(j)})]=0\}=1.\\
\end{equation}
Thus we have $\lim_{N\rightarrow\infty}\Pr\{I(\mathrm{M}^t,\mathrm{E}^t,\mathrm{F}^t;\mathrm{Z}^t)=0\}=1$ and then $\lim_{N\rightarrow\infty}\Pr\{\mathrm{L}(T)=0\}=1$, which proves that strong security can be achieved.
\end{IEEEproof}
\end{proposition}

\begin{proposition}\label{prop_secrecyrate}With with $N\rightarrow\infty$, secrecy capacity of AWTC model can be achieved by the proposed secure polar coding scheme.
\begin{IEEEproof}
Since for entire $T$ blocks, message bits $\mathrm{M}^{1:T}$ are transmitted over subset $\mathcal{I}\setminus\mathcal{E}$, which is proven secure and reliable with infinite $N$, we have the secrecy rate as
\begin{equation}\label{eq_srate}
\begin{split}
\mathrm{R_s}(T)&=\frac{1}{TN}\sum_{t=1}^{T}|\mathcal{I}\setminus\mathcal{E}|\\
&=\frac{1}{N}|\mathcal{I}\setminus\mathcal{E}|.
\end{split}
\end{equation}
Then with $N\rightarrow\infty$, have
\begin{equation}
\begin{split}
\lim_{N\rightarrow\infty}\mathrm{R_s}(T)&=\lim_{N\rightarrow\infty}\frac{1}{N}|\mathcal{I}\setminus\mathcal{E}|\\
&=\lim_{N\rightarrow\infty}\frac{1}{N}(|\mathcal{I}\cup\mathcal{R}|-|\mathcal{E}\cup\mathcal{R}|)\\
&=\lim_{N\rightarrow\infty}\frac{1}{N}(|\mathcal{I}\cup\mathcal{R}|-|\mathcal{B}\cup\mathcal{R}|)\\
&=\lim_{N\rightarrow\infty}\frac{1}{N}(|\mathcal{L}_w|-|\mathcal{H}_r^c|)\\
&\overset{(a)}=1-\rho_w-\rho_r,
\end{split}
\end{equation}
where $(a)$ is due to \eqref{eq_rate} that
\begin{equation}
\begin{split}
\lim_{N\rightarrow\infty}\frac{1}{N}|\mathcal{L}_w|&=1-\rho_w, \\
\lim_{N\rightarrow\infty}\frac{1}{N}|\mathcal{H}_r^c|&=\lim_{N\rightarrow\infty}\frac{1}{N}|\mathcal{L}_r|\\
&=1-(1-\rho_r)\\
&=\rho_r.
\end{split}
\end{equation}
Thus the secrecy capacity of AWTC model can be achieved with with $N\rightarrow\infty$.
\end{IEEEproof}
\end{proposition}

\subsection{Simulations}

\begin{figure}[!t]
\centering
\subfloat[Upper bound of legitimate BER]{\includegraphics[width=8.2cm]{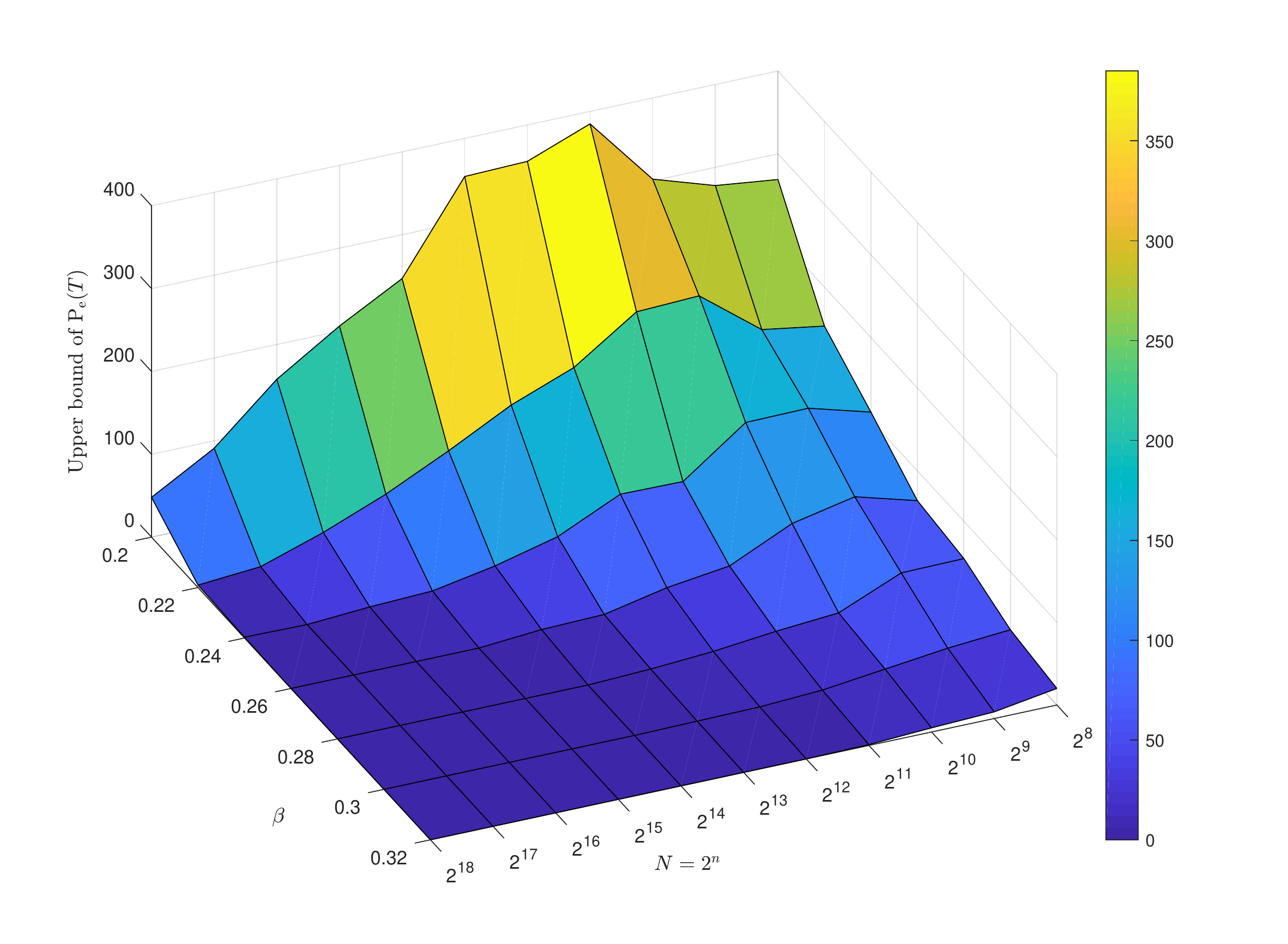}%
\label{fig_ulb}}
\hfil
\subfloat[Upper bound of information leakage]{\includegraphics[width=8.2cm]{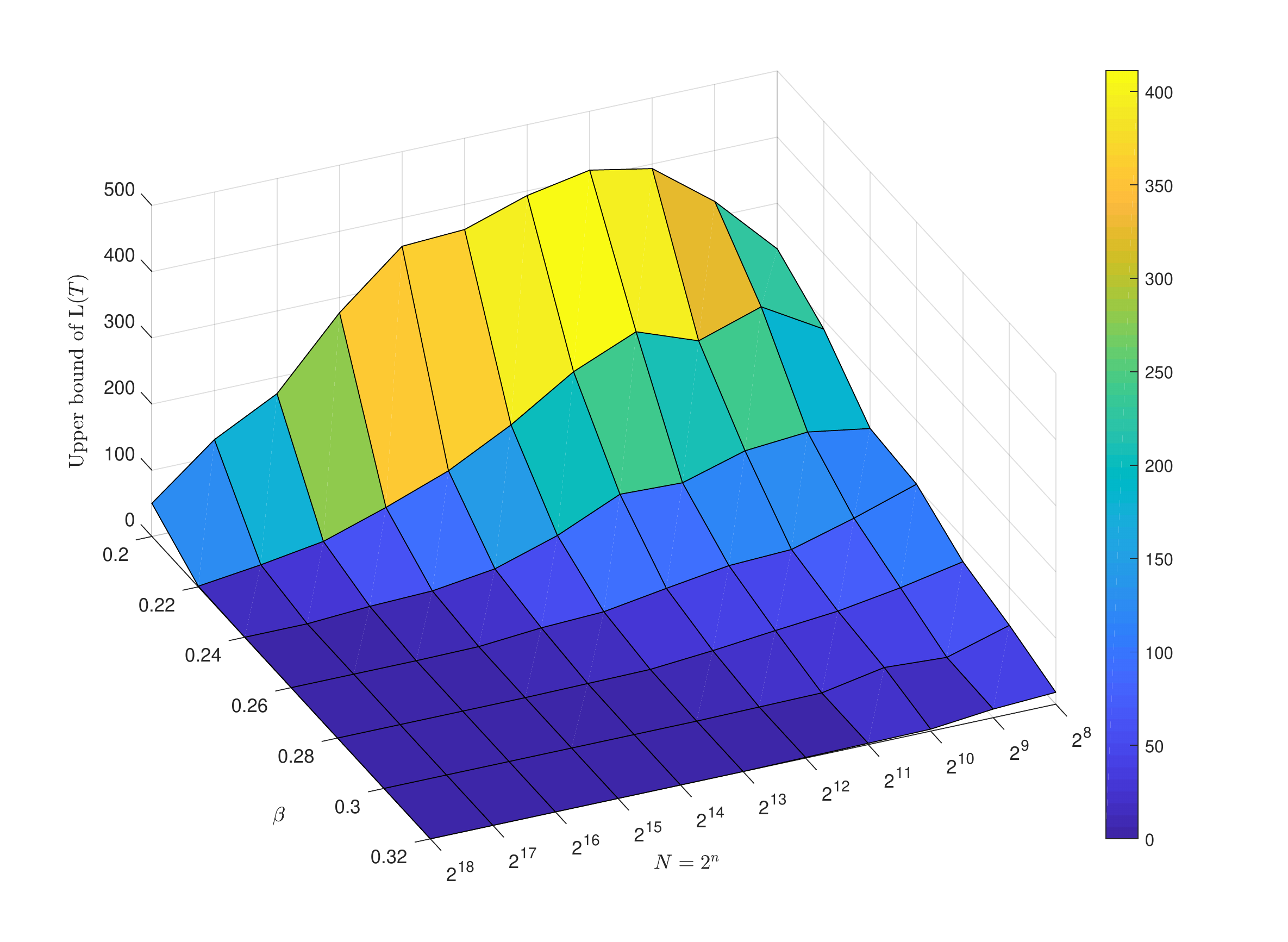}%
\label{fig_uil}}
\hfil
\subfloat[Secrecy rate]{\includegraphics[width=8.2cm]{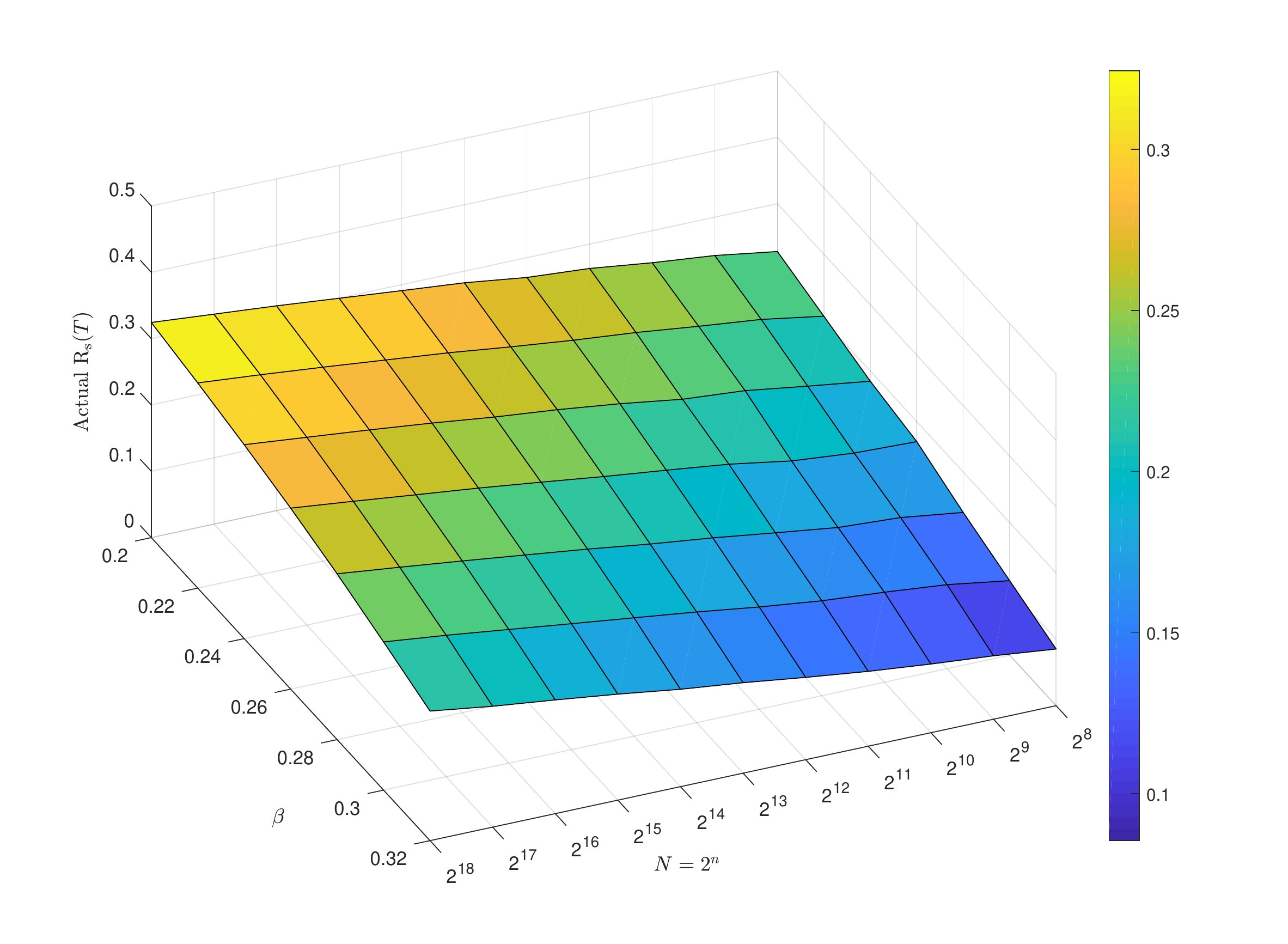}%
\label{fig_sr}}
\caption{Simulation results of performance with finite block length $N$.}
\label{fig_theo_sim}
\end{figure}

Now we test the upper bound of legitimate bit error rate (BER) in \eqref{eq_ber}, the upper bound of information leakage in \eqref{eq_laekage} and the secrecy rate in \eqref{eq_srate} by simulations. Particularly, we fix writhing fraction $\rho_w=0.2$, reading fraction $\rho_r=0.4$ and block number $T=300$, by which we test the performances when block length $N$ changes from $2^8$ to $2^{18}$ and parameter $\beta$ changes from $0.20$ to $0.32$. The simulation results are illustrated in Fig.~\ref{fig_theo_sim}.

\begin{itemize}
  \item Fig.~\ref{fig_ulb} shows that for all $\beta$, the upper bound of legitimate BER keeps going down to $0$ as the block length $N$ increases, which matches our analysis of reliability in Prop.~\ref{prop_reliability}.
  \item Fig.~\ref{fig_uil} shows that for all $\beta$, the upper bound of information leakage keeps going down to $0$ as the block length $N$ increases, which matches our analysis of strong security in Prop.~\ref{prop_security}.
  \item Fig.~\ref{fig_sr} shows for all $\beta$, the actual secrecy rate keeps growing as the block length $N$ increases, which matches our analysis of secrecy rate in Prop.~\ref{prop_secrecyrate}.
\end{itemize}

\begin{figure}[!t]
\centering
\subfloat[Actual BER of Bob]{\includegraphics[width=8cm]{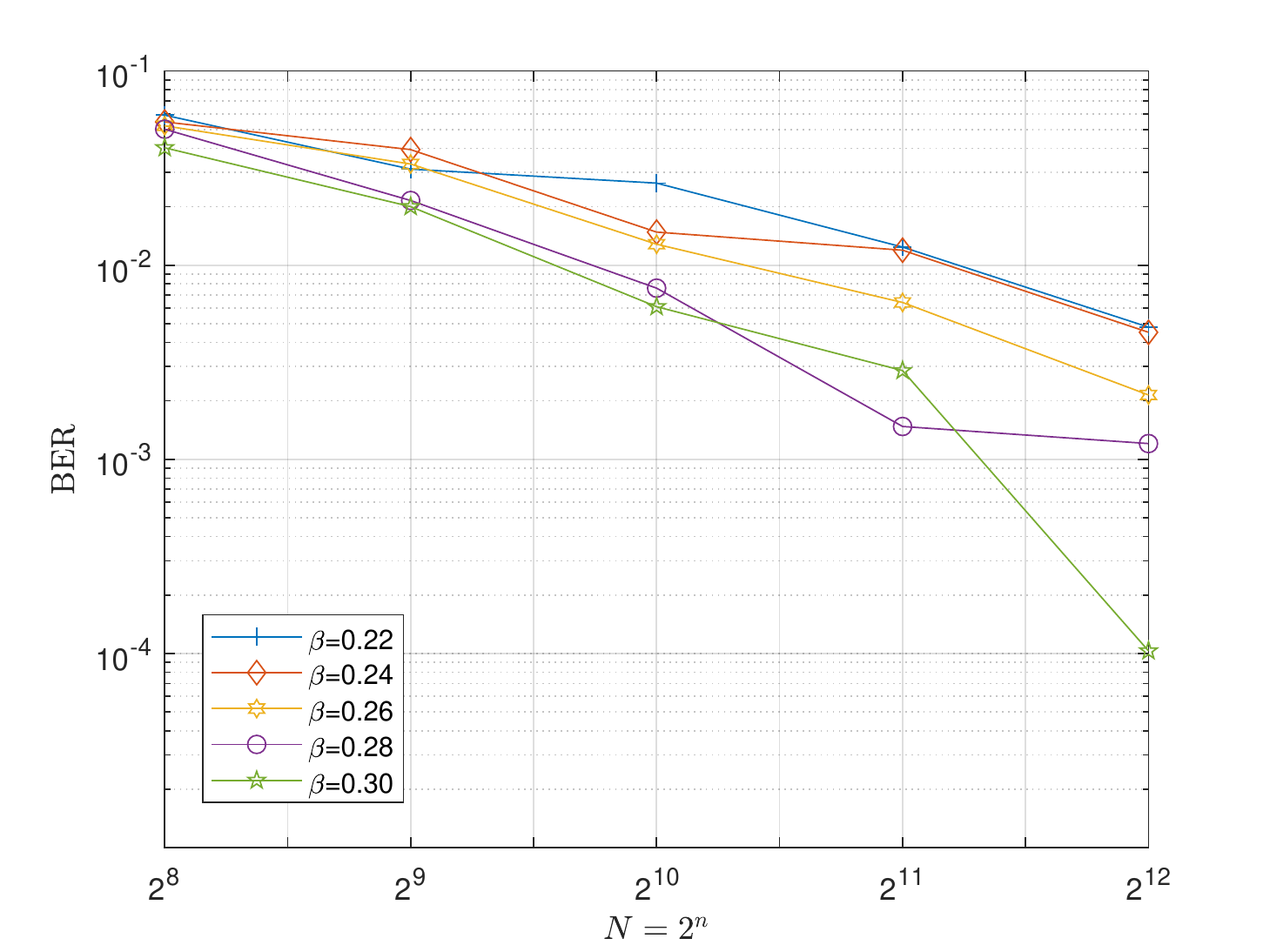}%
\label{fig_ler}}
\hfil
\subfloat[Actual BER of Eve]{\includegraphics[width=8cm]{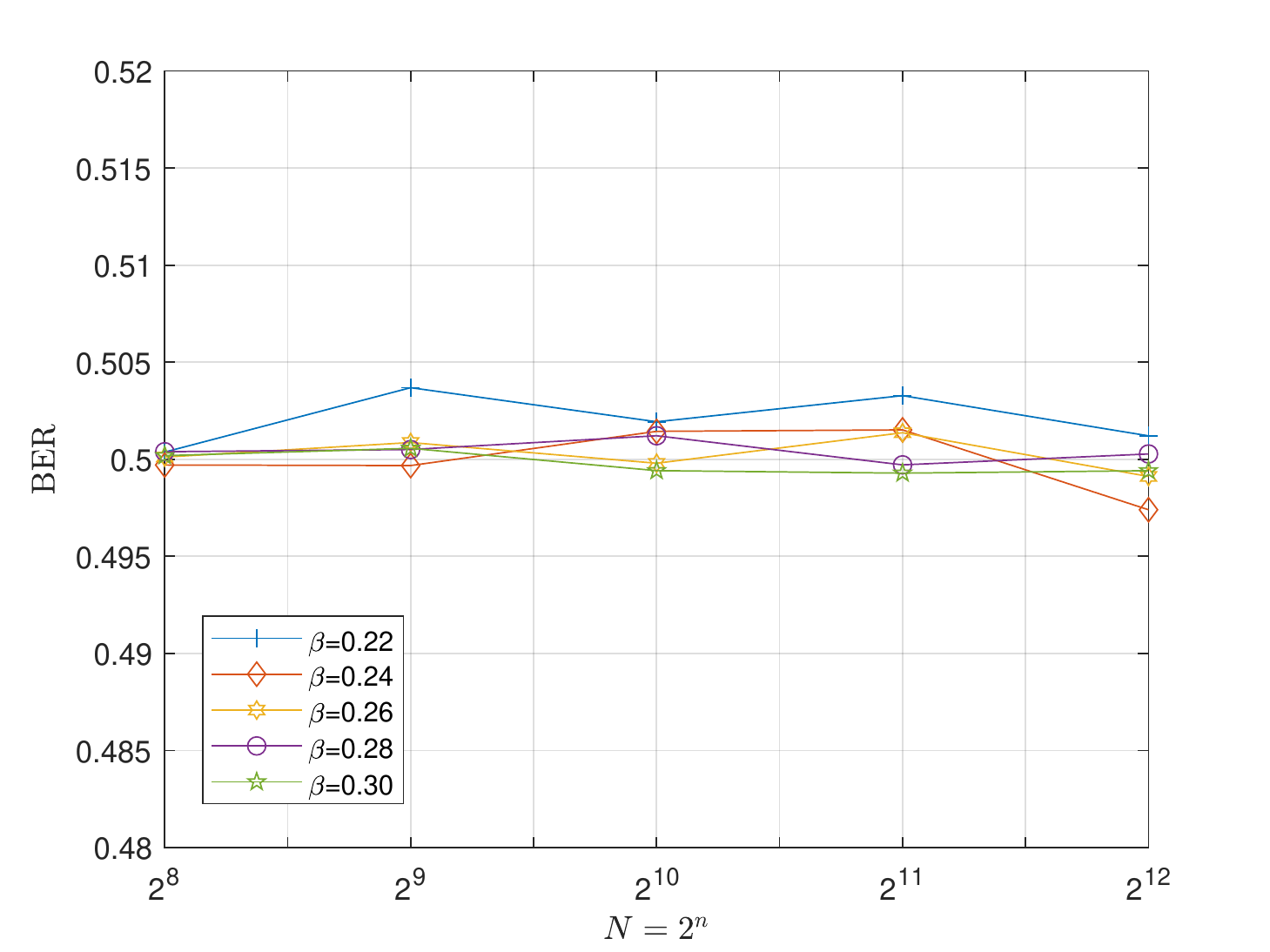}%
\label{fig_eer}}
\caption{Simulation results of actual BERs.}
\label{fig_ber_sim}
\end{figure}

Next, we simulate the entire secure communication process over the AWTC model and test the actual BERs for both Bob and Eve. Particularly, we fix $\rho_w=0.2$, $\rho_r=0.4$ and $T=1000$. Then we consider cases that $N$ goes from $2^8$ to $2^{12}$ and $\beta$ goes from $0.22$ to $0.30$. For the transmitted message, we use uniformly distributed binary random bits. The simulation results are illustrated in Fig.~\ref{fig_ber_sim}.

\begin{itemize}
  \item Fig.~\ref{fig_ler} is the BER of entire $1000$ blocks for legitimate user Bob decoding the message bits, which shows that for all tested $\beta$, the legitimate BER decreases significantly with the increasing of block length $N$.
  \item Fig.~\ref{fig_eer} is the BER of entire $1000$ blocks for eavesdropper Eve decoding the message bits, which shows that for all tested $\beta$, the eavesdropper BER stays closely to $0.5$ with the increasing of block length $N$.
\end{itemize}

Therefore, Fig.~\ref{fig_ber_sim} proves that the proposed securer polar coding scheme can provide secure and reliable communication over the AWTC model.

\section{Conclusion}\label{sec_con}

In this paper, we have considered the secure coding problem of AWTC model. We have presented an channel based equivalent model of AWTC model by using the $\rho$ equivalent channel block. Then we have studied the polarization of the $\rho$ equivalent channel block and proved that it can be polarized by the non-stationary channel polarization operation $\mathbf{G}_N$ in the sense that $1-\rho$ of the generated channels are noiseless with probability almost $1$ and $\rho$ of the generated channels are full-noise with probability almost $1$. Based on this results, for the AWTC model, we polarized both equivalent channel blocks of adversarial reading and writing actions and constructed a sucre polar coding scheme by applying the multi-block chaining structure. Theoretically, we have proven that the proposed scheme achieves the secrecy capacity of AWTC model under both reliability and strong security criterions with an infinite block length $N$. Further we have carried out simulations which proves that the proposed secure polar coding scheme can provide secure and reliable communication over AWTC model.

% or


%
\appendices
%% you can choose not to have a title for an appendix
%% if you want by leaving the argument blank
%\section{}

\section*{Acknowledgment}

This work is supported in part by the National Natural Science Foundation of China (Grand No.62004077) and Natural Science Foundation of Hubei Province (Grant No.2019CFB137).

\ifCLASSOPTIONcaptionsoff
  \newpage
\fi

\begin{IEEEbiographynophoto}{Yizhi Zhao}
received the Ph.D. degree in the school of Optical and Electronic Information from the Huazhong University of Science and Technology, Wuhan,
China, in 2017.

He is currently an Assistant Professor with the College of Informatics, Huazhong Agricultural University. His research interests include physical layer coding, information theory and machine learning.
\end{IEEEbiographynophoto}

%\begin{IEEEbiographynophoto}{Hongmei Chi}
%received her Ph.D. degree in the School of Mathematics and Statistics from Wuhan University, Wuhan, China, in 2014.
%
%Currently, she is an Assistant Professor with College of Science, Huazhong Agricultural University. Her research interest is statistic learning, stochastic analysis and information theory.

\end{document}